\documentclass[journal]{IEEEtran}
\usepackage[english]{babel}
\usepackage{graphicx}
\usepackage{amsmath}
\usepackage{mathtools}
\usepackage{cite}

\begin{document}

\title{Information Leakage of Heterogeneous Encoded Correlated Sequences over Eavesdropped Channel}
\author{Reevana Balmahoon and Ling Cheng
\\
\authorblockA{School of Electrical and Information Engineering\\University of the Witwatersrand\\Private Bag 3, Wits. 2050, Johannesburg, South Africa\\
Email: reevana.balmahoon@students.wits.ac.za, ling.cheng@wits.ac.za } }
\maketitle

\begin{abstract}
Correlated sources are present in communication systems where protocols ensure that there is some predetermined information for sources. Here correlated sources across an eavesdropped channel that incorporate a heterogeneous encoding scheme and their effect on the information leakage when some channel information and a source have been wiretapped is investigated. The information leakage bounds for the Slepian-Wolf scenario are provided. Thereafter, the Shannon cipher system approach is presented. Further, an implementation method using a matrix partition approach is described. 

\end{abstract}

\section{Introduction}

Practical communication systems make use of correlated sources, for example smart grid meters. Each smart grid meter for a particular grid conforms to certain protocols and this means that certain information (e.g. date, area, etc.) in the header files will be the same for various meters. From the receiver's (or an eavesdropper's) perspective, it appears as common information shared between the meters. This is therefore pre-existing or known information for an eavesdropper. Thus, correlated sources are common in systems transmitting information, e.g. smart grid meter systems. This implies that the theory used for correlated sources may also be applied to this type of system. 

Correlated source coding incorporates the lossless compression of two or more correlated data streams. Correlated sources have the ability to decrease the bandwidth required to transmit and receive messages because a compressed form of the original message is sent across the communication links instead of the original message. A compressed message has more information per bit, and therefore has a higher entropy because the transmitted information is more unpredictable. The unpredictability of the compressed message is also beneficial for the information security. 

In practical communication systems links are prone to eavesdropping and as such this work incorporates wiretapped channels, more specifically the wiretap channel II. In work by Aggarwal \textit{et al.} \cite{active_eavesdropper_aggarwal} it is seen that an eavesdropper can be active and can erase/modify bits. They develop a perfect secrecy model for this scenario. The eavesdropper that we investigate is a passive wiretapper, who cannot modify information. The mathematical model for this wiretap channel has been given by Rouayheb \textit{et al.}~\cite{ref12_rouayheb_soljanin}, and can be explained as follows: the channel between a transmitter and receiver is error-free and can transmit $n$ symbols from which $\mu$ of them can be observed by the eavesdropper and the maximum secure rate can be shown to equal $n-\mu$ symbols. The wiretap channel II was described by Ozarow and Wyner \cite{ref15_ozarow_wyner} with a coset coding scheme. This wiretap channel can even be looked at from a Gaussian approach. A variation of this Gaussian wiretap channel has been investigated by Mitrpant \textit{et al.} \cite{gaussian_wiretap_mitrpant}. In this work we use an information theory approach and provide a link to coding theory. There has been work done on wiretap channels for a coding approach. The first was done by Wei \cite{ref13_wei} who presented the generalized Hamming weight to describe the minimum uncertainty that an eavesdropper has access to. Thereafter characteristics on this channel were introduced by Luo \textit{et al.} \cite{ref14_luo_mitpant}. The characteristics focused on were those pertaining to Hamming weights and Hamming distances in order to determine the equivocation of a wiretapper. Thereafter, the security aspect of wiretap networks has been looked at in various ways by Cheng \textit{et al.} \cite{ref21_cheng_yeung}, and Cai and Yeung \cite{ref11_cai_yeung}, emphasizing that it is of concern to secure this type of channels. 

The difference between the original wiretap channel and the wiretap channel II is that the latter is error free. In an interesting application of the wiretap channel and wiretap channel of type II, Dai \textit{et al.} \cite{side_information_dai} presented a model that incorporates compromised encoded bits and wiretapped bits from a noisy channel. The concept of a noiseless transmission gives rise to an ideal situation in terms of noise when analyzing the model. Here, we consider a scenario where an eavesdropper has access to more than just the bits from the communication links. Luo \textit{et al.} \cite{ref14_luo_mitpant}, in some previous work, have described a similar sort of adversary as more powerful. In addition to the eavesdropped bits from the communication links, the eavesdropper also has access to some data symbols from the two remaining sources. In other previous work \cite{arxiv1_bal_ling}, the information leakage for two correlated sources when some channel information from the communication links had been wiretapped was investigated. Intuitively from this work, it is seen that there is indeed more information gained by the more powerful eavesdropper, not just in terms of the source symbols but in terms of the source being considered, which results from the fact that the sources are correlated. This makes it easier for the eavesdropper to determine the transmitted message and information about the source of concern.

This extra information that the eavesdropper has access to can be considered as side information to assist with decoding. Villard and Piantanida \cite{pablo_secure_multiterminal} have also looked at correlated sources and wiretap networks: A source sends information to the receiver and an eavesdropper has access to information correlated to the source, which is used as side information. There is a second encoder that sends a compressed version of its own correlation observation of the source privately to the receiver. Here, the authors show that the use of correlation decreases the required communication rate and increases secrecy. Villard \textit{et al.} \cite{pablo_secure_transmission_receivers} have explored this side information concept further where security using side information at the receiver and eavesdropper is investigated. Side information is generally used to assist the decoder to determine the transmitted message. An earlier work involving side information was done by Yang \textit{et al.}~\cite{feedback_yang}. The concept can be considered to be generalized in that the side information could represent a source. It is an interesting problem when one source is more important and Hayashi and Yamamoto\cite{Hayashi_coding} have considered it in another scheme with two sources, where only one source is secure against wiretappers and the other must be transmitted to a legitimate receiver. They develop a security criterion based on the number of correct guesses of a wiretapper to attain a message. In this paper the source data symbols may be seen as side information to the eavesdropper, which is further explained in Section II.

Shannon's secrecy model is an interesting avenue for this work. Previous work \cite{arxiv1_bal_ling} has looked at a model for Shannon's cipher system when there is wiretapping at the channel only. Merhav \cite{shannon_secrecy_merhav} investigated similarly, for a model using the additional parameters of the distortion of the source reconstruction at the legitimate receiver, the bandwidth expansion factor of the coded channels, and the average transmission cost.

In the model presented herein two correlated sources and a third source having correlation to one other source, which may also be wiretapped is considered. The paper is arranged in nine sections. Section II puts forth a description of the model and Section III presents the information leakage quantification for this model. The information leakage is a new concept developed and quantifies how much of information the adversary/eavesdropper has access to. The proofs for the information leakage quantification are presented in Section IV. In Section V, the Shannon cipher system approach is presented, where channel and key rates for perfect secrecy are determined. In Section VI, the practical investigation is detailed. Thereafter, similar models are discussed in Section VII and the paper is concluded in Section VIII and the Appendix that details the proofs for the theorems developed for the Shannon cipher approach are contained in Section IX. 

\section{Model}
\begin{figure}[ht]
\centering
\includegraphics [scale = 0.7]{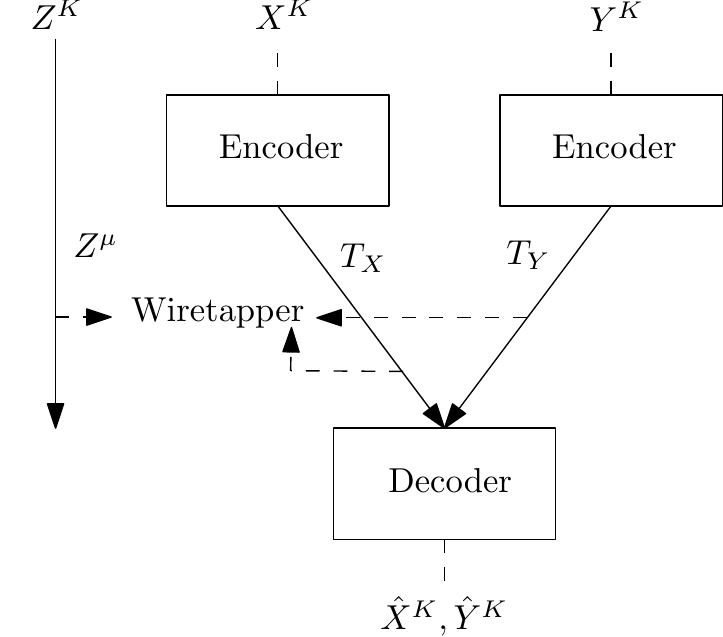}
\caption{Correlated source coding for heterogeneous encoding scheme with $X$ and $Y$ transmitting compressed information}
\label{fig:new_model3}
\end{figure}

\begin{figure}[ht]
\centering
\includegraphics [scale = 0.7]{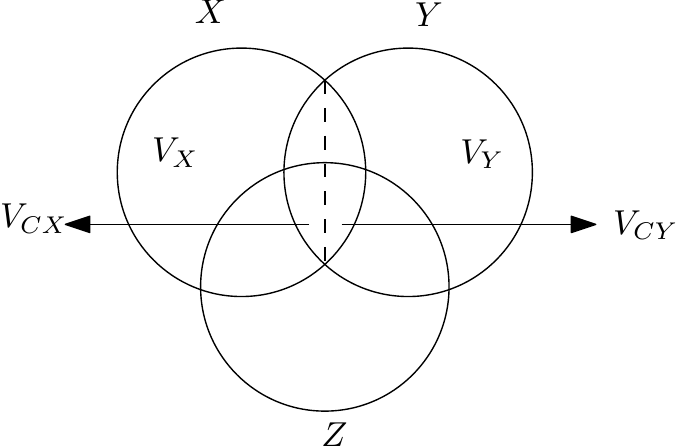}
\caption{The relation between private and common information}
\label{fig:new_venn3}
\end{figure}

The independent, identically distributed (i.i.d.) sources $X$, $Y$ and $Z$ are mutually correlated random variables, depicted in Figure~\ref{fig:new_model3}. The alphabet sets for sources $X$, $Y$ and $Z$ are represented by $\mathcal{X}$ and $\mathcal{Y}$ and $\mathcal{Z}$ respectively. Assume that $X^K$ and $Y^K$ are encoded into two channel information portions represented by their common and private information portions. We can write $T_X = (V_{CX}, V_X)$ and $T_Y = (V_{CY}, V_Y)$ where $T_X$ and $T_Y$ are the channel information of $X$ and $Y$ respectively. The Venn diagram in Figure \ref{fig:new_venn3} easily illustrates this idea. Each source is composed of $K$ bits, and for source $Z^K$, $\mu$ of these symbols are considered as the predetermined information and is leaked to the wiretapper ($\mu \le K$).

This encoding scheme, as specified in \cite{Balmahoon_arXiv} reaches the Slepian-Wolf bound. Here, the length of $T_X$ and $T_Y$ is not fixed, as it depends on the encoding process and nature of the Slepian-Wolf codes.

The correlated sources $X$, $Y$ and $Z$ transmit messages (in the form of some channel information) to the receiver along the channel. The decoder determines $X$ and $Y$ only after receiving $T_X$ and $T_Y$.  

The eavesdropper has access to either the common or private portions given by $T_X$ and $T_Y$ and the eavesdropped source information $Z^{\mu}$. The effect is that the eavesdropper has access to some compressed information (that is transmitted across the communication link after encoding) and some uncompressed information (i.e. the source $Z$'s data symbols). It is valuable to determine how much of information this eavesdropper has access to when wiretapping the private or common information portions (this is described in the next section).

Here, for $X^K$ and $Y^K$ typical set encoding and decoding is used. We are able to determine bin indices for the typical sequence from the indices passed over the communication channel. When common or private information from the syndromes are wiretapped it gives an indication of which row/column in the specific look up table the sequence is contained within. The encoding and decoding for $X$ and $Y$ has been described in detail in previous work \cite{Balmahoon_arXiv}. The decoding probabilities follow. 

From the Venn diagram we see that the private information and common information produced by each source should contain almost no redundancy. 
Here, $V_{CX}$, $V_X$, $V_Y$, $V_{CY}$ are asymptotically disjoint, which ensures that there is almost no redundant information sent to the decoder. 

In previous work \cite{arxiv1_bal_ling}, we have considered the common information that $V_{CX}$ and $V_{CY}$ represent, which was found to be $I(X;Y)$. Here, we begin to explore the nature of codes when there are three correlated sources. We first define the prototype code: For any $\epsilon_0 \geq 0$ and sufficiently large $K$, there exits a code $W_{CX} = F_{CX}(X^K)$, $W_{CY} = F_{CY}(Y^K)$, $\widehat{X}^K,\widehat{Y}^K, \widehat{Z}^K$, where $W_X \in I_{M_X}$, $W_Y \in I_{M_Y}$, $W_{CX} \in I_{M_{CX}}$ and $W_{CY} \in I_{M_{CY}}$ for $I_{M_{\alpha}}$, which is defined as $\{0, 1, \ldots, M_{\alpha} - 1\}$, that satisfies,

\begin{eqnarray}
Pr \{\widehat{X}^K, \widehat{Y}^K \neq X^K, Y^K\} \le \epsilon_0
\label{lemma1_1}
\end{eqnarray}

\begin{eqnarray}
H(X|Y,Z) - \epsilon_0 & \le & \frac{1}{K} H(W_X) \le \frac{1}{K} \log M_X  \nonumber\\
& \le & H(X|Y,Z) + \epsilon_0
\label{lemma1_2}
\end{eqnarray}

\begin{eqnarray}
H(Y|X,Z) - \epsilon_0 & \le & \frac{1}{K} H(W_Y) \le \frac{1}{K} \log M_Y  \nonumber\\
& \le & H(Y|X) + \epsilon_0
\label{lemma1_3}
\end{eqnarray}

\begin{eqnarray}
& & I(X;Y) - \epsilon_0 \le \frac{1}{K} \log [H(W_{CX}) + H(W_{CY})] \nonumber\\
& \le &  I(X;Y) + \epsilon_0
\label{lemma1_4}
\end{eqnarray}

\begin{eqnarray}
\frac{1}{K} H(X^K|V_Y) \geq H(X) - \epsilon_0
\label{lemma1_5}
\end{eqnarray}

\begin{eqnarray}
\frac{1}{K} H(Y^K|V_X) \geq H(Y) - \epsilon_0
\label{lemma1_6}
\end{eqnarray}

\begin{eqnarray}
\frac{1}{K} H(Z^K|V_Y) \geq H(Z) - \epsilon_0
\label{lemma1_8}
\end{eqnarray}

We can see that \eqref{lemma1_1} - \eqref{lemma1_4} mean
\begin{eqnarray}
&& H(X,Y) - 3\epsilon_0 \le \frac{1}{K} (H(W_X) + H(W_{CX}) + H(W_Y) \nonumber \\
& + & H(W_{CY})) \le H(X,Y) + 3\epsilon_0 
\label{lemma1_7}
\end{eqnarray}

Hence from \eqref{lemma1_1}, \eqref{lemma1_7} and the ordinary source coding theorem, ($W_X$, $W_Y$, $W_{CX}$ and $W_{CY}$) have almost no redundancy for sufficiently small $\epsilon_0 \geq 0$. Equations \eqref{lemma1_1} - \eqref{lemma1_7} have been proven in \cite{arxiv1_bal_ling} for two sources.

This model can cater for a scenario where a particular source, say $Y$ needs to be more secure than $X$ (possibly because of eavesdropping on the $Y$ channel only); we would need to secure the information that could be compromised. A masking approach to achieve this is described in previous work \cite{Balmahoon_arXiv}. 

\section{Information Leakage Using Slepian-Wolf Approach}
In order to determine the security of the system, a measure for the amount of information leaked has been developed. The obtained information and total uncertainty are used to determine the leaked information. Information leakage is indicated using $L_{\mathcal{Q}}^\mathcal{P}$. Here $\mathcal{P}$ indicates the set of sources for which information leakage is being quantified. Further, $\mathcal{Q}$ indicates the transmitted sequence that has been wiretapped. 

The information leakage bound for the cases where information from all three links are wiretapped is investigated. There are two cases considered:
\\
Case 1: Leakage on $Y$ when $T_{Y}, T_X, Z^{\mu}$ are wiretapped.
\\
Case 2: Leakage on $X$ when $T_{Y}, T_X, Z^{\mu}$ are wiretapped.
 
The information leakage for these cases is as follows:

\begin{eqnarray}
L_{T_{Y}, T_X, Z^{\mu}}^{Y^K} & \le & \frac{1}{K}[H(V_Y) + H(V_{CY}) - H(Y^K) \nonumber\\
& + & I(T_Y;Y^K) + I(T_X; Y^K) \nonumber\\
& + & I(T_Y;T_X|Y^K) + I(Y^K; Z^{\mu}) + I(T_Y;Z^{\mu}|Y^K)		\nonumber\\
& + & I(T_X;Z^{\mu}|Y^K, T_Y) - I(T_X;T_Y) - I(T_X;Z^{\mu})		\nonumber\\
& - & I(T_Y; Z^{\mu}|T_X)] + \delta
\label{L_inequality1}
\end{eqnarray}

\begin{eqnarray}
L_{T_{Y}, T_X, Z^{\mu}}^{X^K} & \le & \frac{1}{K}[H(V_X) + H(V_{CX}) - H(X^K) \nonumber\\
& + & I(T_Y; X^K) + I(X^K;T_X) \nonumber\\
& + & I(T_Y;T_X|X^K) + I(X^K;Z^{\mu})+ I(T_Y;Z^{\mu}|X^K) 		\nonumber\\
& + & I(T_X; Z^{\mu}|X^K, T_Y) - I(T_X;T_Y) - I(Z^{\mu};T_X)		\nonumber\\
& - & I(T_Y;Z^{\mu}|T_X)] + \delta
\label{L_inequality2}
\end{eqnarray}

Here, $T_X$ and $T_Y$ are the compressed sequences and in terms of the information quantity they include either the private or common portion. Thus, we can see the above bound as a generalised result for wiretapping $X$'s or $Y$'s links, when the source $Z$ is leaked. 
The portion $Z^{\mu}$ could be leaked from the private or common portion of $Z$; this is therefore also a generalised representation for the leaked portion for $Z$.

This is interesting because this case deals with correlated sources and as such intuitively it is known that there is some information that may be leaked by an alternate source. The common information component between all sources is the maximum information that can be leaked by another source. For this case, the common information $V_{CX}$ and $V_{CY}$ can thus consist of added protection to reduce the amount of information leaked. 

\section{Proof of Information Leakage Bounds}
This bound developed in \eqref{L_inequality1} is proven below.
\\\\ \textit{Proof for \eqref{L_inequality1}}: 
First, $H(Y^K|T_Y,T_X,Z^{\mu})$ is determined, as to perform the information leakage calculation we need to find $H(Y) - H(Y|T_Y,T_X, Z^{\mu})$
\begin{eqnarray}
& & \frac{1}{K} H(Y^K|T_Y,T_X,Z^{\mu})											\nonumber
\\
& = & \frac{1}{K} [H(Y^K,T_Y,T_X,Z^{\mu}) - H(T_Y,T_X,Z^{\mu})] 			\nonumber\\
& \stackrel{(a)}{=} & \frac{1}{K} [H(Y^K) + H(T_Y|Y^K) + H(T_X|T_Y, Y^K) 					\nonumber\\
& + & H(Z^{\mu}|Y^K, T_Y, T_X) - (H(T_Y) + H(T_X|T_Y)		\nonumber\\
& + & H(Z^{\mu}|T_Y, T_X))]										\nonumber\\
& \stackrel{(b)}{=} & \frac{1}{K} [H(Y^K) + (H(T_Y) - I(T_Y;Y^K)) + (H(T_X 	\nonumber\\
& - & I(Y^K;T_X)	-I(T_Y; T_X|Y^K)) + (H(Z^{\mu}) - I(Y^K;Z^{\mu})				\nonumber\\
& - & I(T_Y; Z^{\mu}|Y^K) - I(T_X; Z^{\mu}|Y^K,T_Y)) - H(T_Y)		\nonumber\\
& - & (H(T_X) - I(T_X; T_Y)) - (H(Z^{\mu}) 					\nonumber\\
& - & I(Z^{\mu}; T_X) - I(T_Y;Z^{\mu}|T_X))]					\nonumber\\
& \stackrel {(c)}{=} & \frac{1}{K} [H(Y^K) + H(T_Y) - I(T_Y; Y^K) + H(T_X) 	\nonumber\\
& - & I(Y^K;T_X)) - I(T_Y;T_X|Y^K) + H(Z^{\mu}) - I(Y^K;Z^{\mu}) 					\nonumber\\
& - & I(T_Y;Z^{\mu}|Y^K) - I(T_X; Z^{\mu}|Y^K, T_Y) -H(T_Y)			\nonumber\\
& - & H(T_X) + I(T_X;T_Y) - H(Z^{\mu}) + I(Z^{\mu};T_X)	\nonumber\\
& + & I(T_Y;Z^{\mu}|T_X)]							\nonumber\\
& = & \frac{1}{K} [H(Y^K) - I(T_Y; Y^K) - I(Y^K;T_X)	\nonumber\\		
& - & I(T_Y;T_X|Y^K) - I(Y^K;Z^{\mu})- I(T_Y;Z^{\mu}|Y^K) 			\nonumber\\
& - & I(T_X; Z^{\mu}|Y^K, T_Y) + I(T_X;T_Y)  					\nonumber\\
& + & I(Z^{\mu};T_X) +  I(T_Y;Z^{\mu}|T_X)]				
\end{eqnarray}

where $(a)$ results from the chain rule expansion for $ H(Y,T_Y,T_X,Z^{\mu})$ and $H(T_Y,T_X,Z^{\mu})$ and $(b)$ results from the property that the conditional entropy is the same as the mutual information subtracted from the total uncertainty, i.e. $H(X|Y) = H(X) - I(X;Y)$. Here, $(c)$ is arithmetic, where the terms $H(T_Y)$, $ H(T_X)$ and $H(Z^{\mu})$ cancel.

The information leakage is thus:
\begin{eqnarray}
L_{T_Y,T_X,Z^{\mu}}^{Y^K} & = & H(Y) - \frac{1}{K} H(Y^K|T_Y,T_X,Z^{\mu})	\nonumber\\ 
& \le & \frac{1}{K}[H(V_Y) + H(V_{CY}) - H(Y^K) \nonumber\\ 
& + & I(T_Y; Y^K) + I(Y^K;T_X) + I(T_Y;T_X|Y^K) 	\nonumber\\		
& + & I(Y^K;Z^{\mu})+ I(T_Y;Z^{\mu}|Y^K) 			\nonumber\\
& + & I(T_X; Z^{\mu}|Y^K, T_Y) - (I(T_X;T_Y)  					\nonumber\\
& + & I(Z^{\mu};T_X) +  I(T_Y;Z^{\mu}|T_X))] - \delta				\nonumber\\
& = & \frac{1}{K}[H(V_Y) + H(V_{CY}) - H(Y^K) 			\nonumber\\
& + & I(T_Y; Y^K) + I(Y^K;T_X)+ I(T_Y;T_X|Y^K) 					\nonumber\\
& + & I(Y^K;Z^{\mu})+ I(T_Y;Z^{\mu}|Y^K) 			\nonumber\\
& + & I(T_X; Z^{\mu}|Y^K, T_Y) - I(T_X;T_Y) 		\nonumber\\
& - & I(Z^{\mu};T_X) - I(T_Y;Z^{\mu}|T_X)] + \delta 		
\label{Lemma2_proof_inequality1}
\end{eqnarray}
which proves \eqref{L_inequality1}.
\\\\ \textit{Proof for \eqref{L_inequality2}}: 
\begin{eqnarray}
& & \frac{1}{K}H(X^K|T_Y,T_X,Z^{\mu})											\nonumber
\\
& = & \frac{1}{K}[H(X^K,T_Y,T_X,Z^{\mu}) - H(T_Y,T_X,Z^{\mu})] 			\nonumber\\
& \stackrel{(d)}{=} & \frac{1}{K}[H(X^K) + H(T_Y|X^K) + H(T_X|T_Y, X^K) 					\nonumber\\
& + & H(Z^{\mu}|X^K, T_Y, T_X) - (H(T_Y) + H(T_X|T_Y)		\nonumber\\
& + & H(Z^{\mu}|T_Y, T_X))]										\nonumber\\
& \stackrel{(e)}{=} & \frac{1}{K}[H(X^K) + (H(T_Y) - I(T_Y;X^K)) 	\nonumber\\
& + & (H(T_X - I(X^K;T_X)- I(T_Y; T_X|X^K)) + (H(Z^{\mu}) 				\nonumber\\
& - & I(X^K;Z^{\mu}) - I(T_Y; Z^{\mu}|X^K) - I(T_X; Z^{\mu}|X^K,T_Y)) 		\nonumber\\
& - & H(T_Y) - (H(T_X) - I(T_X; T_Y)) - (H(Z^{\mu}) 					\nonumber\\
& - & I(Z^{\mu}; T_X) - I(T_Y;Z^{\mu}|T_X))]					\nonumber\\
& \stackrel {(f)}{=} & \frac{1}{K}[H(X) + H(T_Y) - I(T_Y; X^K) + H(T_X) 	\nonumber\\
& - & I(X^K;T_X)) - I(T_Y;T_X|X^K) + H(Z^{\mu}) - I(X^K;Z^{\mu}) 					\nonumber\\
& - & I(T_Y;Z^{\mu}|X^K) - I(T_X; Z^{\mu}|X^K, T_Y) -H(T_Y)			\nonumber\\
& - & H(T_X) + I(T_X;T_Y) - H(Z^{\mu}) + I(Z^{\mu};T_X)	\nonumber\\
& + & I(T_Y;Z^{\mu}|T_X)]										\nonumber\\
& = & \frac{1}{K}[H(X^K) - I(T_Y; X^K) - I(X^K;T_X)	\nonumber\\		
& - & I(T_Y;T_X|X^K) - I(X^K;Z^{\mu})- I(T_Y;Z^{\mu}|X^K) 			\nonumber\\
& - & I(T_X; Z^{\mu}|X^K, T_Y) + I(T_X;T_Y)  					\nonumber\\
& + & I(Z^{\mu};T_X) +  I(T_Y;Z^{\mu}|T_X)]				
\end{eqnarray}

where $(d)$ results from the chain rule expansion for $ H(X,T_Y,T_X,Z^{\mu})$ and $H(T_Y,T_X,Z^{\mu})$ and $(e)$ results from the property that the conditional entropy is the same as the mutual information subtracted from the total uncertainty, i.e. $H(X|Y) = H(X) - I(X;Y)$. Here, $(f)$ is arithmetic, where the terms $H(T_Y)$, $ H(T_X)$ and $H(Z^{\mu})$ cancel.

The information leakage is thus:
\begin{eqnarray}
L_{T_Y,T_X,Z^{\mu}}^{X} & = & H(X^K) - \frac{1}{K} H(X^K|T_Y,T_X,Z^{\mu})	\nonumber\\ 
& \le & \frac{1}{K}[H(V_X) + H(V_{CX})- H(X^K) \nonumber\\
& + & I(T_Y; X^K) + I(X^K;T_X)	\nonumber\\		
& + & I(T_Y;T_X|X^K) + I(X^K;Z^{\mu})+ I(T_Y;Z^{\mu}|X^K) 			\nonumber\\
& + & I(T_X; Z^{\mu}|X^K, T_Y) - (I(T_X;T_Y)  					\nonumber\\
& + & I(Z^{\mu};T_X) +  I(T_Y;Z^{\mu}|T_X))] - \delta				\nonumber\\
& = & \frac{1}{K}[I(T_Y; X^K) + I(X^K;T_X) \nonumber\\
& + & I(T_Y;T_X|X^K) + I(X^K;Z^{\mu}) 					\nonumber\\
& + & + I(T_Y;Z^{\mu}|X^K) + I(T_X; Z^{\mu}|X^K, T_Y) 		\nonumber\\
& - & I(T_X;T_Y) - I(Z^{\mu};T_X) 							\nonumber\\
& - &  I(T_Y;Z^{\mu}|T_X)] + \delta	
\label{Lemma2_proof_inequality2}
\end{eqnarray}
which proves \eqref{L_inequality2}.

This section shows the information leakage for when various portions of the channel information and some source data symbols are leaked. It is evident that the eavesdropper has more information about a particular source as shown in \eqref{Lemma2_proof_inequality1} and \eqref{Lemma2_proof_inequality2} than if only one or two links transmitting compressed information were wiretapped. This can be drawn from a comparison of some previous work \cite{arxiv1_bal_ling}. The interesting cases explored for \eqref{L_inequality1} and \eqref{L_inequality2} demonstrate that the source $Z$ contributes to leakage for $X$ and $Y$; this is due to the common information shared between them. 

Equations \eqref{L_inequality1} and \eqref{L_inequality2} indicate that the information leakage is upper bounded by the common information portions indicated. The information leakage in \eqref{L_inequality1} and \eqref{L_inequality2} can be reduced if the common information portions are secured. Equations \eqref{L_inequality1} and \eqref{L_inequality2} can be verified using the Venn diagram in Figure \ref{fig:new_venn3}.

\section{Shannon Cipher Approach}

\begin{figure}[ht]
\centering
\includegraphics [scale = 0.7]{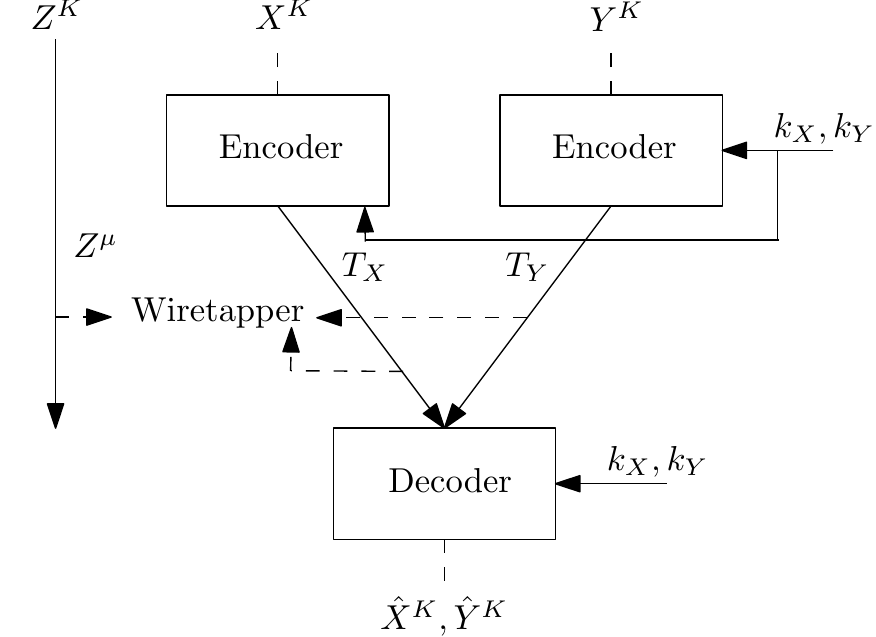}
\caption{Shannon cipher system for three correlated sources}
\label{fig:shannon_cipher_3sources}
\end{figure}

Here, we discuss Shannon's cipher system for three correlated sources (depicted in Figure \ref{fig:shannon_cipher_3sources}). The two source outputs are i.i.d random variables $X$ and $Y$, taking on values in the finite sets $\mathcal{X}$ and $\mathcal{Y}$. Both the transmitter and receiver have access to the key, a random variable, independent of $X^K$ and $Y^K$ and taking values in $I_{M_k} = \{0, 1, 2, \ldots ,M_{k} - 1\}$. The sources $X^K$ and $Y^K$ compute the ciphertexts $W_1$ and $W_2$, which are the result of specific encryption functions on the plaintext from $X$ and $Y$ respectively. The encryption functions are invertible, thus knowing $W_1$ and the key, $k_X$ for $X$ then $X$ can be retrieved. The key for $Y$ is represented as $k_Y$. 

The mutual information between the plaintext and ciphertext should be small so that the wiretapper cannot gain much information about the plaintext. For perfect secrecy, this mutual information should be zero, then the length of the key should be at least the length of the plaintext.

The encoder functions for $X$ and $Y$, ($E_X$ and $E_Y$ respectively) are given as:

\begin{eqnarray}
E_X : \mathcal{X}^K \times I_{M_{kX}} & \rightarrow & I_{M_X'} =  \{0, 1, \ldots, M_X' - 1\} \nonumber 
\\ && I_{M_{CX}'} =  \{0, 1, \ldots, M_{CX}' - 1\}
\label{xencoder_fcn}
\end{eqnarray}

\begin{eqnarray}
E_Y : \mathcal{Y}^K \times I_{M_{kY}} & \rightarrow & I_{M_Y'} =  \{0, 1, \ldots, M_Y' - 1\} \nonumber 
\\ && I_{M_{CY}'} =  \{0, 1, \ldots, M_{CY}' - 1\}
\label{yencoder_fcn}
\end{eqnarray}

The decoder is defined as:

\begin{eqnarray}
D_{XY} : (I_{M'_X}, I_{M'_Y}, I_{M'_{CX}},I_{M'_{CY}})  & \times &  I_{M_{kX}}, I_{M_{kY}} \nonumber \\
& \rightarrow & \mathcal{X}^K \times \mathcal{Y}^K
\end{eqnarray}

The encoder and decoder mappings are below:
\begin{eqnarray}
W_1 = F_{E_X} (X^K, W_{kX})
\end{eqnarray}

\begin{eqnarray}
W_2 = F_{E_Y} (Y^K, W_{kY})
\end{eqnarray}

\begin{eqnarray}
\widehat{X}^K = F_{D_X} (W_1, W_2, W_{kX})
\end{eqnarray}

\begin{eqnarray}
\widehat{Y}^K = F_{D_Y} (W_1, W_2, W_{kY})
\end{eqnarray}

or 

\begin{eqnarray}
(\widehat{X}^K, \widehat{Y}^K) = F_{D_{XY}} (W_1, W_2, W_{kX}, W_{kY})
\end{eqnarray}

The following conditions should be satisfied for cases 1- 4:

\begin{eqnarray}
\frac{1}{K}\log M_X \le R_X +\epsilon
\label{cond1}
\end{eqnarray}

\begin{eqnarray}
\frac{1}{K}\log M_Y \le R_Y +\epsilon
\label{cond2}
\end{eqnarray}

\begin{eqnarray}
\frac{1}{K}\log M_{kX} \le R_{kX} +\epsilon
\label{cond3}
\end{eqnarray}

\begin{eqnarray}
\frac{1}{K}\log M_{kY} \le R_{{kY}} +\epsilon
\label{cond4}
\end{eqnarray}

\begin{eqnarray}
\text {Pr} \{\widehat{X}^K \neq X^K\} \le \epsilon
\label{cond5}
\end{eqnarray}

\begin{eqnarray}
\text{Pr} \{ \widehat{Y}^K \neq Y^K\} \le \epsilon
\label{cond6}
\end{eqnarray}

\begin{eqnarray}
\frac{1}{K} H(X^K|W_1) \le h_X + \epsilon
\label{cond7}
\end{eqnarray}

\begin{eqnarray}
\frac{1}{K} H(Y^K|W_2) \le h_Y + \epsilon
\label{cond8}
\end{eqnarray}

\begin{eqnarray}
\frac{1}{K} H(X^K,Y^K|W_1, W_2) \le h_{XY} + \epsilon
\label{cond8.1}
\end{eqnarray}

where $R_X$ is the rate of source $X$'s channel and $R_Y$ is the rate of source $Y$'s channel. Here, $(R_{kX}, R_{kY})$ is the rate of the key channel when allocating a key to $X$ and $Y$. The security level for $X$ and $Y$ are measured by the total and individual uncertainties, $(h_X, h_Y)$ and $h_{XY}$ respectively. 
\\\\
The cases 1 - 3 that are considered are as follows:
\\ \textit{Case 1:} When $(W_1, W_2, Z^{\mu})$ is leaked and $(X^K, Y^K)$ needs to be kept secret. The security level of concern is represented by $\frac{1}{K}H(X^K, Y^K|W_1, W_2, Z^{\mu})$.
\\ \textit{Case 2:} When $(W_1, W_2, Z^{\mu})$ is leaked and $(X^K, Y^K)$ needs to be kept secret. The security level of concern is represented by $(\frac{1}{K}H(X^K|W_1, W_2, Z^{\mu}), \frac{1}{K}H(Y^K|W_1, W_2, Z^{\mu}))$.
\\ \textit{Case 3:} When $(W_1, W_2, Z^{\mu})$ is leaked and $Y^K$ needs to be kept secret.The security level of concern is represented by $\frac{1}{K}H(Y^K|W_1, W_2, Z^{\mu})$.
\\ 
\\\\
The admissible rate region for each case is defined as follows:
\\ \textit{Definition 1a:} ($R_X$, $R_Y$, $R_{kX}$, $R_{kY}$, $h_{XY}$) is admissible for case 1 if there exists a code ($F_{E_{X}}$, $F_{D_{XY}}$) and ($F_{E_{Y}}$, $F_{D_{XY}}$) such that \eqref{cond1} - \eqref{cond6} and \eqref{cond8.1} hold for any $\epsilon \rightarrow 0$ and sufficiently large $K$.
\\ \textit{Definition 1b:} ($R_X$, $R_Y$, $R_{kX}$, $R_{kY}$, $h_{X}$, $h_{Y}$) is admissible for case 2 if there exists a code ($F_{E_{X}}$, $F_{D_{XY}}$) and ($F_{E_{Y}}$, $F_{D_{XY}}$) such that \eqref{cond1} - \eqref{cond8} hold for any $\epsilon \rightarrow 0$ and sufficiently large $K$.
\\ \textit{Definition 1c:} ($R_X$, $R_Y$, $R_{kX}$, $R_{kY}$, $h_{Y}$) is admissible for case 3 if there exists a code ($F_{E_{Y}}$, $F_{D_{XY}}$) such that \eqref{cond1} - \eqref{cond6} and \eqref{cond8} hold for any $\epsilon \rightarrow 0$ and sufficiently large $K$.
\\ \textit{Definition 2:}  The admissible rate regions of $\mathcal{R}_j$ for case $j$ are defined as:

\begin{eqnarray}
\mathcal{R}_1(h_{XY}) = \{(R_X, R_Y, R_{kX}, R_{kY}):			\nonumber
\\(R_X, R_Y, R_{kX}, R_{kY}, h_{XY} ) \text{ is admissible for case 1} \}
\end{eqnarray}

\begin{eqnarray}
\mathcal{R}_2(h_X, h_{Y}) = \{(R_X, R_Y, R_{kX}, R_{kY}):			\nonumber
\\ (R_X, R_Y, R_{kX}, R_{kY},  h_{X}, h_{Y} ) \text{ is admissible for case 2} \}
\end{eqnarray}

\begin{eqnarray}
\mathcal{R}_3(h_{Y}) = \{(R_X, R_Y, R_{kX}, R_{kY}):			\nonumber
\\ (R_X, R_Y, R_{kX}, R_{kY}, h_{Y} ) \text{ is admissible for case 3} \}
\end{eqnarray}

Theorems for these regions have been developed:

\textit{Theorem 1:} For $0 \le h_{XY} \le H(X,Y) - \alpha_{CX} - \alpha_{CY} + I(X;Y;Z)$,
\begin{eqnarray}
&& \mathcal{R}_1(h_{XY}) = \{(R_X, R_Y, R_{kX},R_{kY}): 		\nonumber
\\ && R_X \geq H(X|Y), 				\nonumber
\\ && R_Y \geq H(Y|X),		\nonumber
\\ && R_X + R_Y	\geq H(X,Y)						\nonumber
\\ && R_{kX} +  R_{kY} \geq h_{XY} \}			
\label{theorem2}
\end{eqnarray}

\textit{Theorem 2:} For $0 \le h_{X} \le H(X) - \alpha_{CX}$  and 
\\ $0  \le h_{Y} \le H(Y) - \alpha_{CY}$
\begin{eqnarray}
&& \mathcal{R}_2(h_{Y}) = \{(R_X, R_Y, R_{kX},R_{kY}): 		\nonumber
\\ && R_X \geq H(X|Y), 				\nonumber
\\ && R_Y \geq H(Y|X),		\nonumber
\\ && R_X + R_Y	\geq H(X,Y)						\nonumber
\\ && R_{kX} + R_{kY} \geq \text{max}(h_X, h_Y) \}			
\label{theorem3}
\end{eqnarray}

where $\mathcal{R}_{1}$ and $\mathcal{R}_{2}$ are the regions for cases 1 and 2 respectively. Here, $\alpha_{CX}$ and $\alpha_{CY}$ are the common portions (i.e. the correlated information) of the i.i.d source $Z$ (for $I(X;Z)$ and $I(Y;Z)$ respectively) that are contained in $Z^{\mu}$ per symbol. 
When $h_X = 0$ then case $3$ can be reduced to that depicted in \eqref{theorem3}.  
Hence, Corollary 1 follows:
\\ \textit{Corollary 1:} For $0 \le h_Y \le H(Y) - \alpha_{CY}$,
\\$\mathcal{R}_{3} (h_Y) = \mathcal{R}_{2}(0, h_Y)$

The security levels, which are measured by the total and individual uncertainties $h_{XY}$ and $(h_X, h_Y)$ respectively give an indication of the level of uncertainty in knowing certain information. When the uncertainty increases then less information is known to an eavesdropper and there is a higher level of security. The proofs of Theorems 1 and 2 are detailed in the appendix.

\section{Information Leakage Using Matrix Partitions}
In this section the aim is to determine the equivocation (uncertainty) in retrieving the message from the transmitted channel information. We follow the convention used by Stankovic \textit{et al.} \cite{ref10_stankovic_liveris} to present an example together with a method incorporating generator matrix ranks put forth by Luo \textit{et al.} \cite{ref14_luo_mitpant} and determine the equivocation. The Hamming distances are represented as follows: $d_H (X^K,Y^K) \le 1$ and $d_H (Y^K,Z^K) \le 1$. It is noted that there is some sort of correlation between $X$ and $Z$ due to the Hamming distance relations between $(X^K, Y^K)$ and $(Y^K,Z^K)$. 

The following generator matrix $G$ is used:
\[
G =
  \begin{bmatrix}
    1 & 0 & 0 & 0 & 1 & 0 & 1\\
    0 & 1 & 0 & 0 & 1 & 1 & 0\\
    0 & 0 & 1 & 0 & 1 & 1 & 1\\
    0 & 0 & 0 & 1 & 0 & 1 & 1\\
  \end{bmatrix}
\]

The matrix takes the form: $G = [I_k P^T]$ and here $I_k$ is the identity matrix of order $k$ and $P^T$ is made up of two $2 \times 3$ matrices in this case. 
 
Suppose the messages to send across the channels for $X$ and $Y$ are given by: 
$x = [a_1\text{ } v_1\text{ } q_1] = [10\text{ } 11\text{ } 001]$ and $y = [u_2 \text{ } a_2 \text{ } q_2] = [10\text{ } 11\text{ } 011]$.

There is compression along $X$'s and $Y$'s channel. As per the matrix partition method the syndrome for $X$ and $Y$ is comprised of:

\[
T_X =
  \begin{bmatrix}
    v_1^T\\
    P_1^Ta_1^T \oplus q_1^T\\
  \end{bmatrix}
\]

\[
T_Y =
  \begin{bmatrix}
    u_2^T\\
    P_2^Ta_2^T \oplus q_2^T\\
  \end{bmatrix}
\]
where $P_1^T$ is the $G$ matrix transpose of rows 1-2 and columns 5-7 and $P_2^T$ is the $G$ matrix transpose of rows 3-4 and columns 5-7.
The generator matrices used by $X$ and $Y$ to achieve these syndromes are $G_X$ and $G_Y$ respectively. 
\[
G_X =
  \begin{bmatrix}
    0 & 0 & 1 & 0 & 1\\
    0 & 0 & 1 & 1 & 0\\
    1 & 0 & 0 & 0 & 0\\
    0 & 1 & 0 & 0 & 0\\
    0 & 0 & 1 & 0 & 0\\
    0 & 0 & 0 & 1 & 0\\
    0 & 0 & 0 & 0 & 1\\
  \end{bmatrix}
\]

\[
G_Y =
  \begin{bmatrix}
    1 & 0 & 0 & 0 & 0\\
    0 & 1 & 0 & 0 & 0\\
    0 & 0 & 1 & 1 & 1\\
    0 & 0 & 0 & 1 & 1\\
    0 & 0 & 1 & 0 & 0\\
    0 & 0 & 0 & 1 & 0\\
    0 & 0 & 0 & 0 & 1\\
  \end{bmatrix}
\]

This results in syndromes of $[1\text{ } 1\text{ } 1\text{ } 0\text{ } 0]$ for $X$ and $[1\text{ } 0\text{ } 1\text{ } 1\text{ } 1]$ for $Y$.

Here, the equivocation for these cases can be found using the $G$ matrix specified above and a sub matrix of $G$. As per Luo \textit{et al.} \cite{ref14_luo_mitpant}, the equivocation is given by: $\triangle_{Y|T_Y} = \text{rank} (G) - \text{rank} (G_Y)$, where $\triangle_{Y|T_Y}$ is the equivocation on $Y$ given $T_Y$.

Next, the information leakage for each of the following cases is analyzed:
\begin{itemize}
\item
The equivocation on $(X^K,Y^K)$ when $(T_X,T_Y)$ is leaked
\item
The equivocation on $(X^K,Y^K)$ when $T_X$ is leaked
\item
The equivocation on $(X^K,Y^K)$ when $T_Y$ is leaked
\end{itemize}

In order to show the most representative results for each of the cases the scenarios contributing to the minimum and maximum information leakage have been considered. 

Before the information leakage method is described certain variables are introduced. Here, $\mu_{T_X}$ and $\mu_{T_Y}$ represent the number of wiretapped bits from $T_X$ and $T_Y$ respectively. The length of the information bits from each syndrome is represented as $l_i^X$ and $l_i^Y$ for $X^K$ and $Y^K$ respectively. The length of parity bits with respect to $X^K$ or $Y^K$ is denoted as $l_p$, and the following can be developed: $l_i^X + l_i^Y + 2l_p$ is the overall length of $T_X$ and $T_Y$. Hence we have: $0 \le \mu_{T_X} \le l_i^X + l_p$ and $0 \le \mu_{T_Y} \le l_i^Y + l_p$.

Note that the leakage is determined using a combination of the information and parity bits and the parity matrix $H$ rank. The $H$ matrix rank is used to determine how much of information is leaked from the wiretapped bits, when the columns corresponding to the wiretapped bits have been removed. Let $H'$ denote the $H$ matrix with the wiretapped columns removed. 

\begin{figure}[ht]
\includegraphics [scale = 0.7] {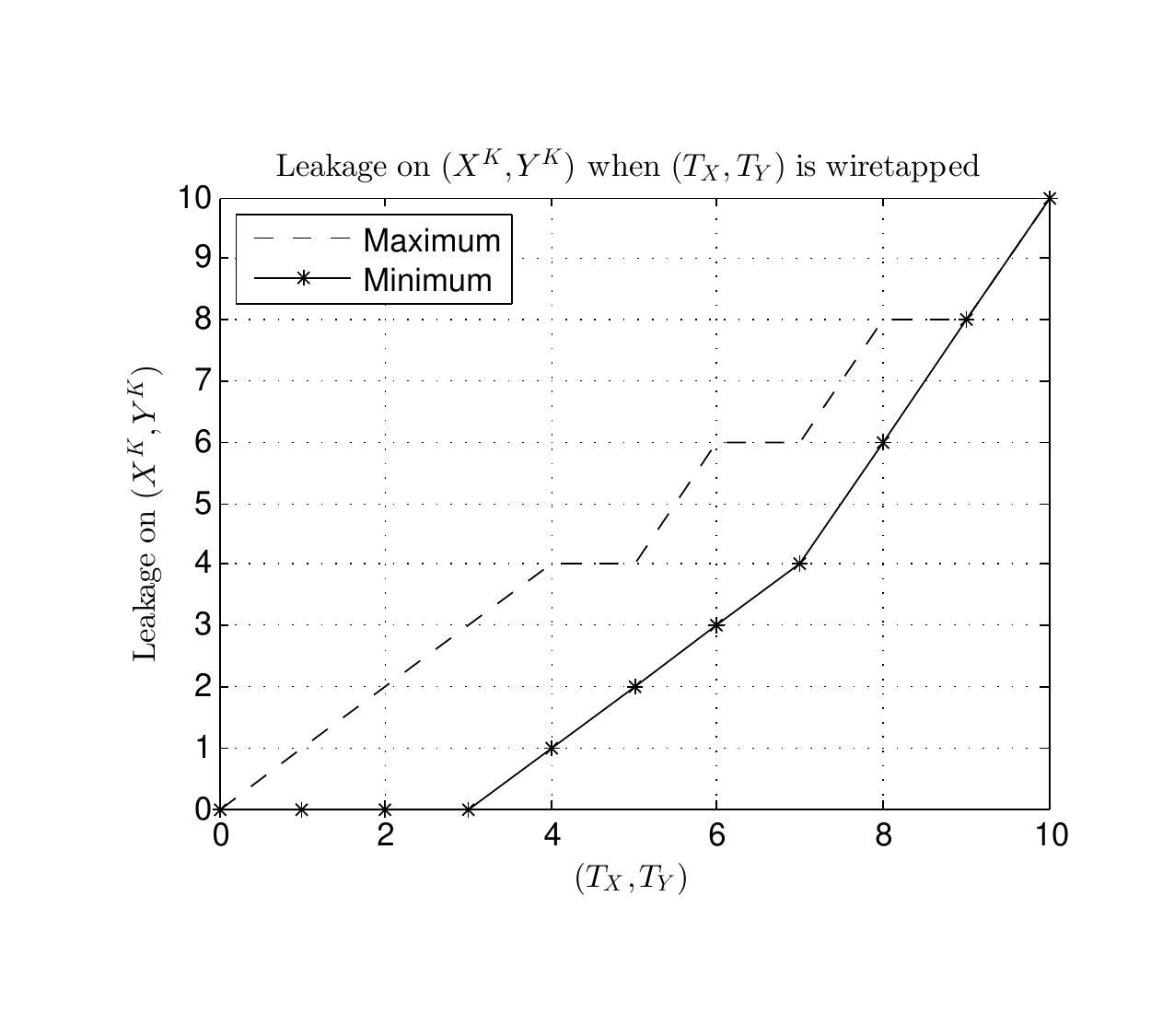}
\caption{The information leakage on $(X^K,Y^K)$ when $(T_X,T_Y)$ has been wiretapped}
\label{fig:txandty}
\end{figure}

The case for the leakage on $(X^K,Y^K)$ when $(T_X, T_Y)$ is leaked is now considered. Initially we consider when the maximum leakage is reached. When $\mu_{T_X} \le l_i^X$ and $\mu_{T_Y} \le l_i^Y$, the maximum leakage is $\mu_{T_X}+ \mu_{T_Y}+ \text{rank}(H)-\text{rank}(H')$. This considers when the information bits (namely $v_1$ and $u_2$) have been leaked only. Next, if $\mu_{T_X} > l_i^X$ and $\mu_{T_Y} > l_i^Y$ is considered. For this case  $\text{min}(\mu_{T_X} - l_i^X, \mu_{T_Y}-l_i^Y)$ parity bits can be from the corresponding positions in $P_1^Ta_1^T \oplus q_1^T$ and $P_2^Ta_2^T \oplus q_2^T$. Therefore, the maximum leakage is $l_i^X + l_i^Y + \mu_{T_X}+ \text{min}(\mu_{T_X} - l_i^X, \mu_{T_Y}-l_i^Y)+\text{rank}(H)-\text{rank}(H')$. If $\mu_{T_X} > l_i^X$ and $\mu_{T_Y} \leq l_i^Y$ the maximum leakage is $\mu_{T_Y}+l_i^X +\text{rank}(H)-\text{rank}(H')$; if $\mu_{T_X} \leq l_i^X$ and $\mu_{T_Y} > l_i^Y$, the maximum leakage is $\mu_{T_X}+l_i^Y +\text{rank}(H)-\text{rank}(H')$. Now we consider when the minimum leakage is reached. When $\mu_{T_X} \le l_p$ and $\mu_{T_Y} \le l_p$, the minimum leakage is $\max(0, \mu_{T_X}+\mu_{T_Y}-l_p)+\text{rank}(H)-\text{rank}(H')$. This considers when the parity bits (namely $P_1^Ta_1^T \oplus q_1^T$ and $P_2^Ta_2^T \oplus q_2^T$) have been leaked only. Otherwise, the minimum leakage is $\mu_{T_X}+\mu_{T_Y}-l_p+\text{rank}(H)-\text{rank}(H')$. For the numerical example considered the information leakage as represented above is depicted in Figure \ref{fig:txandty}.

%  in the  are from  is determined. If $\mu_{T_Y}$ is the minimum there is insufficient information to determine if the parity bits leak any information. If $\mu_{T_X} - l_i^X$ is the minimum there is sufficient information. 
%
%For the first case the maximum leakage is given by \eqref{max_l}.
%\begin{eqnarray}
%L_{T_{X}, T_Y}^{X^K, Y^K} = l_i^X + l_i^Y + \lfloor {\frac{l_p}{2}} \rfloor + (3 - \text{rank} <H'>)
%\label{max_l}
%\end{eqnarray}
%
%where $H'$ is the $H$ matrix with the wiretapped columns removed. 
% 
%For the first case the minimum leakage is given by \eqref{min_l} and \eqref{min_2} for $2 \le l_i^X + l_i^Y \le 4$ and $l_i^X + l_i^Y \ge 5$ respectively. For the case $l_i^X + l_i^Y = 1$ the information leakage on $(X^K,Y^K)$ when $(T_{X}, T_Y)$ is wiretapped is given by $\lfloor {\frac{l_p}{2}} \rfloor + (3 - \text{rank} <H'>)$.
%\begin{eqnarray}
%L_{T_{X}, T_Y}^{X^K, Y^K} = \lfloor {\frac{l_p}{2}} \rfloor - 1 + (3 - \text{rank} <H'>)
%\label{min_l}
%\end{eqnarray}
%
%\begin{eqnarray}
%L_{T_{X}, T_Y}^{X^K, Y^K} = \lfloor {\frac{l_p}{2}} \rfloor + l_i^X + l_i^Y + (3 - \text{rank} <H'>)
%\label{min_2}
%\end{eqnarray}
%
%For the minimum case the columns representing the parity bits are initially removed from the $H$ matrix and in the maximum case the columns representing the information bits are initially removed from the $H$ matrix. 
%
%For the numerical example considered the information leakage as represented in \eqref{max_l}-\eqref{min_2} is depicted in Figure \ref{fig:txandty}.

%L_{Z^{\mu}}^{X^K, Y^K} & = & H(X^K, Y^Y) \nonumber \\
%&& + (K-\mu+1)\frac{K-\mu+1}{2^{K-\mu}(K+1)} \log_2\frac{K-\mu+1}{2^{K-\mu}(K+1)} \nonumber \\
%&& + \mu 2^{K-\mu}\frac{1}{\mu 2^{K-\mu}}\log_2\frac{1}{\mu 2^{K-\mu}} -  H(X^K| Y^Y)

\begin{figure}[ht]
\includegraphics [scale = 0.7] {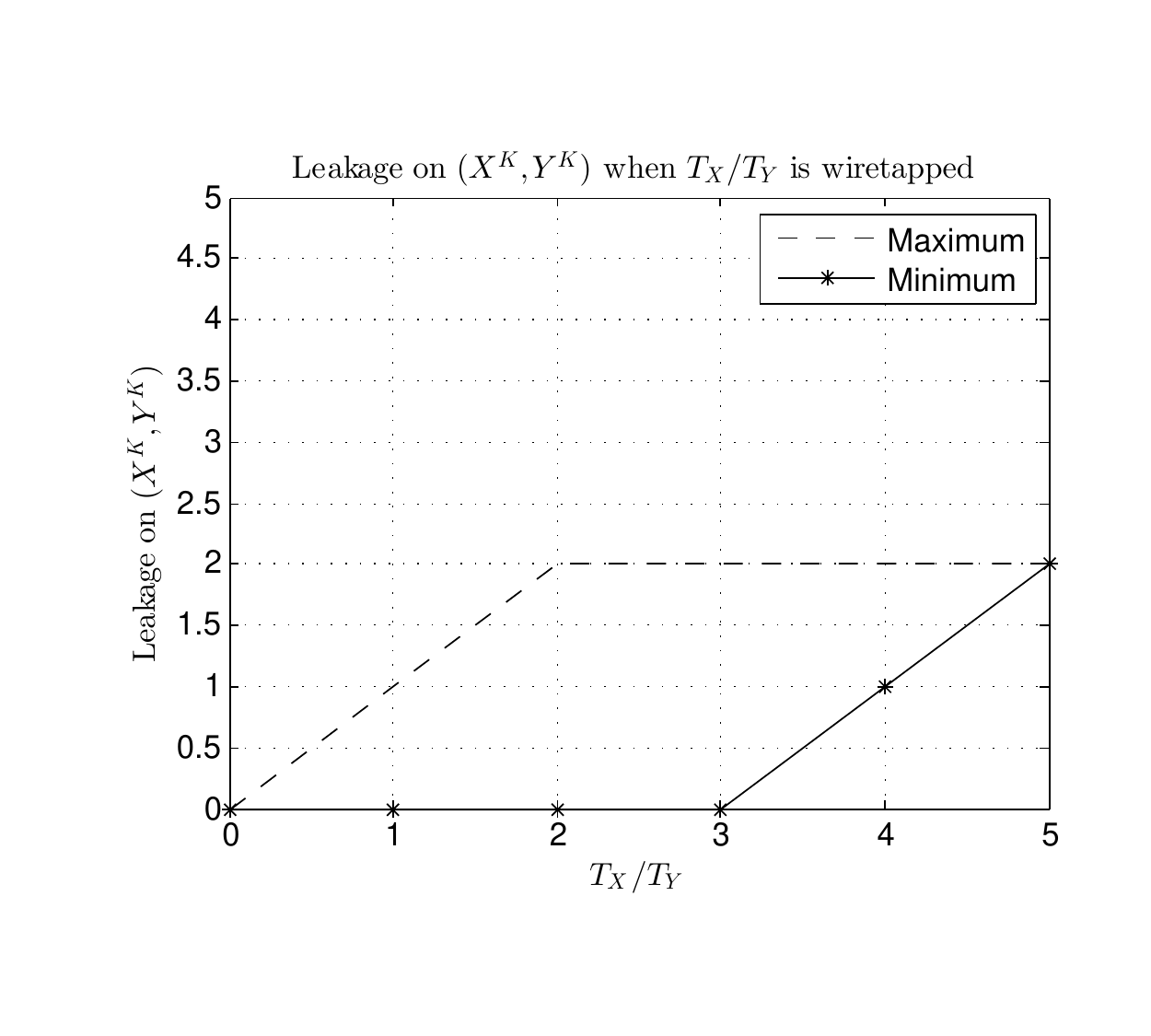}
\caption{The information leakage on $(X^K,Y^K)$ when $T_X$ or $T_Y$ has been wiretapped}
\label{fig:txorty}
\end{figure}

Next the information leakage for the second and third cases are determined. The leakage for these cases reach the same limit for the minimum and maximum cases of information leakage, however the information leakage peak occurs at different points depending on which bits (information or parity) are leaked first. The leakage for the second and third cases respectively are as follows: $L_{T_{X}}^{X^K, Y^K} = l_i^X$ and $L_{T_Y}^{X^K, Y^K} = l_i^Y$. Using the numerical example for this section, the graphical representation is in Figure \ref{fig:txorty}. The maximum case depicted is when the information bits are initially wiretapped and the minimum case is where the parity bits are initially wiretapped. 

Now the information leakage on $(X^K, Y^K)$ when $\mu$ bits of source $Z$ is leaked is considered, which is done in two steps. First, since $d_{H}(Y^K, Z^K) \leq 1$, if there are $0 < \mu \leq K$ bits ($\mu=0$ is considered earlier in this section), the number of possible sequences including repeated sequences is $2^{K-\mu} (K+1)$. However, there are $2^{K-\mu}$ different sequences repeated $K-\mu+1$ times and there are $\mu 2^{K-\mu}$ different sequences that possibly occur once. Second, from every possible $Y^K$, there are eight possible sequences for $X^K$ with identical possibilities. Therefore, the information leakage due to $Z^{\mu}$ with respect to $X^K$ and $Y^K$ is detailed in \eqref{for_z}.

\begin{eqnarray}
&L_{Z^{\mu}}^{X^K, Y^K} = H(X^K, Y^K) + \frac{K-\mu+1}{K+1} \log_2\frac{K-\mu+1}{2^{K-\mu}(K+1)} \nonumber \\
& + \mu 2^{K-\mu}\frac{1}{2^{K-\mu}(K+1)}\log_2\frac{1}{2^{K-\mu}(K+1)} -  H(X^K| Y^K) 
\label{for_z}
\end{eqnarray}
where $H(X^K, Y^K)=10$, $H(X^K| Y^K)=3$ and $K=7$. 

The information leakage represented in \eqref{for_z} may be used in the cases explored in this section to separately determine the information leakage of $Z^{\mu}$ on $X^K$ and $Y^K$.

In this example certain bits have more equivocation than others and as such which bits are wiretapped plays a role in making the system vulnerable at different times. For instance, following from the third case if only $T_Y$ is wiretapped from the parity bits then for the first $3$ bits there is no information leakage due to $T_Y$ as the wiretapper would have encountered the masked bits. The information leakage occurs after the third bit, when $u_2$ is wiretapped. This therefore shows an upper and lower bound on the uncertainty, where the upper bound is given when bits $u_2$ is leaked first and the lower bound is given when the masked portion is first leaked. The parity bits are masked and are thus more difficult to be leaked to an adversary. Parity bits from both sources need to be wiretapped and in the same positions in order to leak information. In addition the $Z^{\mu}$ bits also contribute to the information leakage and the correlation between $X^K, Y^K$ and $Z^{\mu}$ plays a role in determining the overall leakage. 

In general, for a systematic code the columns that have a weight of one would contribute one bit to the information leakage entirely. With use of the matrix partition approach, if the parity bits of both $T_X$ and $T_Y$ are wiretapped (and these bits are from the same columns in each generator matrix) then for every two parity bits wiretapped there is one bit of information leaked. The parity bits and the information bits can also be used to solve the parity matrix to determine the information leakage. If the wiretapped parity bits do not belong to the same columns then there is no information leakage at that point.

This section shows the equivocation for the model set forth in Section II when various portions of the channel information and some source data symbols from $Z$ are leaked.

\section{Discussion}

The model presented herein is a more generalised approach of Yamamoto's~\cite{shannon1_yamamoto} model. If we were to combine the communication links for $X$ and $Y$ into one link and wiretap from the link only, we would have a similar situation as per Yamamoto's ~\cite{shannon1_yamamoto}. 
The information transmitted along the channels do not have a fixed length as per Yamamoto's ~\cite{shannon1_yamamoto} method. Here, the channel information length may vary depending on the encoding procedure and nature of Slepian-Wolf codes, which is another feature of this model. 

At first glance, Yamamoto's model may seem to be a generalisation of the Luo \textit{et al.} \cite{ref14_luo_mitpant} model, however Luo \textit{et al.} \cite{ref14_luo_mitpant} incorporate a wiretapper at the source that introduces a more powerful adversary. In our work, we use a similar concept in that the information known to the eavesdropper is $\mu$ source symbols for $Z$, which makes our model further different to Yamamoto's and Luo \textit{et al.} \cite{ref14_luo_mitpant}. 
 
The work by Yang \textit{et al.}~\cite{feedback_yang} uses the concept of side information to assist the decoder in determining the transmitted message. The side information could be considered to be a source and is related to this work when the side information is considered as correlated information or when side information assists in decoding. Similar work with side information that incorporates wiretappers, by Villard and Piantanida \cite{pablo_secure_multiterminal} and Villard \textit{et al.} \cite{pablo_secure_transmission_receivers} may be generalized in the sense that side information can be considered to be a source, however this new model is distinguishable as channel information transmitted across an error free channel can all be wiretapped. Another point is that the $Z^{\mu}$ wiretapped bits that the eavesdropper has access to helps with reducing the uncertainty of determining a message (as mentioned in Section I), can be considered as side information to the eavesdropper.

\section{Conclusion}
Knowing which components contribute most to information leakage aids in keeping the system more secure, as the required terms can be additionally secured. Here, we analyze the effect of an eavesdropper accessing a source and channel information in terms of the information leakage. In Section III the information leakage for the three correlated source model with the more powerful adversary was quantified and proven. It is seen that the common information portion and the wiretapped source are the weaknesses when information is leaked to the eavesdropper and will need to be secured for decreasing the information leakage. The Shannon cipher system approach has provided channel and key rates for perfect secrecy. A method for practical implementation has also been presented using a matrix partition approach.

\section{Appendix}
This section initially proves the direct parts of Theorems 1 - 2 and thereafter the converse parts.

\subsection{Direct parts}
All the channel rates in the theorems above are in accordance with Slepian-Wolf's theorem, hence there is no need to prove them. 
We construct a code based on the prototype code ($W_X, W_Y, W_{CX}, W_{CY}$) in Lemma 1. In order to include a key in the prototype code, $W_X$ is divided into two parts as per the method used by Yamamoto \cite{shannon1_yamamoto}:
\begin{eqnarray}
W_{X1} = W_X \text{ mod } M_{X1} \in I_{M_{X1}} = \{0, 1, 2, \ldots, M_{X1} - 1\}
\label{theorems2-4_eq_1}
\end{eqnarray}

\begin{eqnarray}
W_{X2} = \frac{W_X - W_{X1}}{M_{X1}} \in I_{M_{X2}} = \{0, 1, 2, \ldots, M_{X2} - 1\}
\label{theorems2-4_eq_2}
\end{eqnarray}

where $M_{X1}$ is a given integer and $M_{X2}$ is the ceiling of $M_X/M_{X1}$. The $M_X/M_{X1}$ is considered an integer for simplicity, because the difference between the ceiling value and the actual value can be ignored when $K$ is sufficiently large. In the same way, $W_Y$ is divided:

\begin{eqnarray}
W_{Y1} = W_Y \text{ mod } M_{Y1} \in I_{M_{Y1}} = \{0, 1, 2, \ldots, M_{Y1} - 1\}
\label{theorems2-4_eq_3}
\end{eqnarray}

\begin{eqnarray}
W_{Y2} = \frac{W_Y - W_{Y1}}{M_{Y1}} \in I_{M_{Y2}} = \{0, 1, 2, \ldots, M_{Y2} - 1\}
\label{theorems2-4_eq_4}
\end{eqnarray}

The common information components $W_{CX}$ and $W_{CY}$ are already portions and are not divided further. In this scenario $W_{CX}+W_{CY}$ lies between $0$ and $I(X;Y)$. It can be represented by $X$ and $Y$, $X$ only or $Y$ only.
It can be shown that when some of the codewords are wiretapped the uncertainties of $X^K$ and $Y^K$ are bounded as follows:

\begin{eqnarray}
\frac{1}{K} H(X^K|W_{X2},W_Y) \geq I(X;Y) + \frac{1}{K} \log M_{X1} - \epsilon_{0}^{'}
\label{theorems2-4_ineq_1}
\end{eqnarray}

\begin{eqnarray}
\frac{1}{K} H(Y^K|W_{X},W_{Y2}) \geq I(X;Y) + \frac{1}{K} \log M_{Y1} - \epsilon_{0}^{'}
\label{theorems2-4_ineq_2}
\end{eqnarray}

\begin{eqnarray}
\frac{1}{K} H(X^K|W_{X},W_{Y2}) \geq I(X;Y) - \epsilon_{0}^{'}
\label{theorems2-4_ineq_3}
\end{eqnarray}

\begin{eqnarray}
\frac{1}{K} H(X^K|W_{X},W_Y, W_{CY}) \geq \frac{1}{K} \log M_{CX} - \epsilon_{0}^{'}
\label{theorems2-4_ineq_4}
\end{eqnarray}

\begin{eqnarray}
\frac{1}{K} H(Y^K|W_{X},W_Y, W_{CY}) \geq \frac{1}{K} \log M_{CX} - \epsilon_{0}^{'}
\label{theorems2-4_ineq_5}
\end{eqnarray}

\begin{eqnarray}
\frac{1}{K} H(X^K|W_Y, W_{CY}) \geq H(X|Y) + \frac{1}{K} \log M_{CX} - \epsilon_{0}^{'}
\label{theorems2-4_ineq_6}
\end{eqnarray}

\begin{eqnarray}
\frac{1}{K} H(Y^K|W_Y, W_{CY}) \geq \frac{1}{K} \log M_{CX} - \epsilon_{0}^{'}
\label{theorems2-4_ineq_7}
\end{eqnarray}

\begin{eqnarray}
\frac{1}{\mu} H(Z^{\mu}) \geq \alpha_Z + \alpha_{CX} + \alpha_{CY} - I(X;Y;Z) - \epsilon_{0}^{'}
\label{theorems2-4_ineq_7.1}
\end{eqnarray}

where $\epsilon_{0}^{'} \rightarrow 0$ as  $\epsilon_{0} \rightarrow 0$.
Here, we indicate the wiretapped source symbols with the entropy in \eqref{theorems2-4_ineq_7.1}, where $\alpha_{CX}$ and $\alpha_{CY}$ are the correlated portions of the i.i.d source $Z$ that are contained in $Z^{\mu}$ per symbol and $\alpha_Z$ is the private portion of $Z$. 

The proofs for \eqref{theorems2-4_ineq_1} - \eqref{theorems2-4_ineq_7} are the same as per Yamamoto's\cite{shannon1_yamamoto} proof in Lemma A1. The difference is that $W_{CX}$, $W_{CY}$, $M_{CX}$ and $M_{CY}$ are described as $W_{C1}$, $W_{C2}$, $M_{C1}$ and $M_{C2}$ respectively by Yamamoto. Here, we consider that $W_{CX}$ and $W_{CY}$ are represented by Yamamoto's $W_{C1}$ and $W_{C2}$ respectively. In addition there are some more inequalities considered here:
\begin{eqnarray}
 \frac{1}{K} H(Y^K|W_X, W_{CX}, W_{CY}, W_{Y2}) & \geq & \frac{1}{K} \log M_{Y1}  \nonumber
 \\ & - & \epsilon_{0}^{'}
\label{theorems2-4_ineq_8}
\end{eqnarray}

\begin{eqnarray}
 \frac{1}{K} H(Y^K|W_X, W_{CX}, W_{CY}) & \geq & \frac{1}{K} \log M_{Y1}  \nonumber
\\ & + & \frac{1}{K} \log M_{Y2} - \epsilon_{0}^{'}
\label{theorems2-4_ineq_9}
\end{eqnarray}

\begin{eqnarray}
\frac{1}{K} H(X^K|W_{X2}, W_{CY}) & \geq & \frac{1}{K} \log M_{X1} 	\nonumber
\\ & + & \frac{1}{K} \log M_{CX} - \epsilon_{0}^{'}
\label{theorems2-4_ineq_10}
\end{eqnarray}

\begin{eqnarray}
\frac{1}{K} H(Y^K|W_{X2}, W_{CY}) & \geq & \frac{1}{K} \log M_{Y1} 	\nonumber
\\ & + & \frac{1}{K} \log M_{Y2} + \frac{1}{K} \log M_{CX} 			\nonumber
\\ & - & \epsilon_{0}^{'}
\label{theorems2-4_ineq_11}
\end{eqnarray}

The inequalities \eqref{theorems2-4_ineq_8} and \eqref{theorems2-4_ineq_9} can be proved in the same way as per Yamamoto's\cite{shannon1_yamamoto} Lemma A2, and  \eqref{theorems2-4_ineq_10} and \eqref{theorems2-4_ineq_11} can be proved in the same way as per Yamamoto's\cite{shannon1_yamamoto} Lemma A1.

%%%%%%%%%%%%%%%%%%%%%%%%%%%%%%%%%%%%%%%%%%%%%%%%%%%%%%%%%%%%%%%%%%%%%%%%%%%%%%%%%%%%%%%%%%%%

\begin{proof}[Proof of Theorem 1]
Suppose that ($R_X$, $R_Y$, $R_{KX}$, $R_{KY}$) $\in$ 
$\mathcal{R}_1$ for $h_{XY} \le H(X,Y) - \alpha_{CX} - \alpha_{CY} + I(X;Y;Z)$. Then, from \eqref{theorem2} 
\begin{eqnarray}
&& R_X \geq H(X^K|Y^K)  				\nonumber
\\&& R_Y  \geq H(Y^K|X^K) 				\nonumber
\\&& R_X + R_Y  \geq H(X^K, Y^K)
\label{theorem2_proof_1}
\end{eqnarray}

\begin{eqnarray}
R_{kX} + R_{kY} \geq h_{XY} 
\label{theorem2_proof_2}
\end{eqnarray}

Here the keys are uniform random numbers. For the first case, consider the following: $h_{XY} > I(X;Y)$.

\begin{eqnarray}
M_{X1} = \text{min}(2^{K H(X|Y)}, 2^{K (h_{XY} - I(X;Y))})
\label{theorem2_proof_4.4}
\end{eqnarray}

\begin{eqnarray}
M_{Y1} = 2^{K (h_{XY} - I(X;Y))}
\label{theorem2_proof_4.6}
\end{eqnarray}

The codewords $W_1$ and $W_2$ and the key $W_{kX}$and $W_{kY}$ are now defined:

\begin{eqnarray}
W_1 = (W_{X1} \oplus W_{kY1},  W_{X2}, W_{CX} \oplus W_{kCX}) 
\label{theorem2_proof_7}
\end{eqnarray}

\begin{eqnarray}
W_2 = (W_{Y1} \oplus W_{kY1}, W_{Y2}, W_{CY} \oplus W_{kCY})
\label{theorem2_proof_8}
\end{eqnarray}

\begin{eqnarray}
W_{kX} = W_{kCX}
\label{theorem2_proof_9}
\end{eqnarray}

\begin{eqnarray}
W_{kY} = (W_{kY1}, W_{kCY})
\label{theorem2_proof_10}
\end{eqnarray}

where $W_\alpha \in I_{M_\alpha} = \{0, 1, \ldots, M_\alpha - 1\}$. The wiretapper will not know $W_{X1}$, $W_{CX}$ $W_{Y1}$ and $W_{CY}$ as these are protected by keys.

In this case, $R_X$, $R_Y$, $R_{kX}$ and $R_{kY}$ satisfy from \eqref{theorem2_proof_1} - \eqref{theorem2_proof_10}, that

\begin{eqnarray}
\frac{1}{K} \log M_X + \frac{1}{K} \log M_Y  & = & \frac{1}{K} (\log M_{X1} + \log M_{X2}  \nonumber
\\ & + &  \log M_{CX}) + \frac{1}{K} (\log M_{Y1}  \nonumber
\\ & + &  \log M_{Y2} +  \log M_{CY}) \nonumber 
\\ & \le & H(X|Y) + H(Y|X) 				\nonumber
\\ & + & I(X;Y) + 3 \epsilon_0 \nonumber
\\ & = & H(X,Y) + 3 \epsilon_0 		\nonumber
\\ & \le & R_X + R_Y + 3 \epsilon_0
\label{theorem2_proof_11}
\end{eqnarray}

\begin{eqnarray}
&& \frac{1}{K} [\log M_{kX} + \log M_{kY}]  \nonumber
\\ & = & \frac{1}{K} [\log M_{CX} + \log M_{CY} + \log M_{Y1}] 	\nonumber	
\\ & \le & I(X;Y) + h_{XY} - I(X;Y) -\epsilon_0  \label{num3.2.1}
\\ & = & h_{XY} - \epsilon_0		\nonumber
\\ & \le & R_{kX} + R_{kY} - \epsilon_0
\label{theorem2_proof_13}
\end{eqnarray}

where \eqref{num3.2.1} results from \eqref{theorem2_proof_4.6}.

The security levels thus result:
\begin{eqnarray} 
&& \frac{1}{K} H(X^K, Y^K|W_1, W_2, Z^{\mu}) \nonumber
\\ & = & \frac{1}{K} H(X^K, Y^K|W_{X1} \oplus W_{kY1}, \nonumber
\\ && W_{X2}, W_{CX} \oplus W_{kCX}		\nonumber
\\ && W_{Y1} \oplus W_{kY1}, W_{Y2} \nonumber
\\ && W_{CY} \oplus W_{kCY}, Z^{\mu})			\nonumber
\\ & \ge & \frac{1}{K} H(X^K, Y^K|W_{X1}, W_{X2}, \nonumber
\\ && W_{Y1} \oplus W_{kY1}, W_{Y2}) - \epsilon_0^{''}			\label{num5}
\\ & = & \frac{1}{K} H(X^K, Y^K|W_{X}, W_{Y2}, Z^{\mu} ) - \epsilon_0^{''}		\nonumber
\\ & \geq & I(X;Y) + \frac{1}{K} \log M_{Y1} \nonumber
\\ & - & \alpha_{CX} - \alpha_{CY} + I(X;Y;Z) - 2\epsilon_0^{'} - \epsilon_0^{''}		\nonumber
\\ & = & I(X;Y)+ h_{XY} - I(X;Y)  - \alpha_{CX} - \alpha_{CY} + I(X;Y;Z)  \nonumber
\\ & - & 2\epsilon_0^{'} - \epsilon_0^{''}		\nonumber
\\ & = & h_{XY} - \alpha_{CX} - \alpha_{CY} + I(X;Y;Z)- 2\epsilon_0^{'} - \epsilon_0^{''}
\label{theorem2_proof_16}
\end{eqnarray}

where \eqref{num5} holds because $W_{CX}$ and $W_{CY}$ are secured by uniform random keys and the result of Yamamoto's Lemma A2.

Therefore ($R_X$, $R_Y$, $R_{kX}$, $R_{kY}$, $h_{XY}$) is admissible from \eqref{theorem2_proof_11} -  \eqref{theorem2_proof_16}.

Next the case where: $h_{XY} \le I(X;Y)$ is considered. 
The codewords and keys are now defined:

\begin{eqnarray}
W_1 = (W_{X1},  W_{X2}, W_{CX} \oplus W_{kCX}) 
\label{theorem2_proof_7.1.1}
\end{eqnarray}

\begin{eqnarray}
W_2 = (W_{Y1}, W_{Y2}, W_{CY})
\label{theorem2_proof_8.1.1}
\end{eqnarray}

\begin{eqnarray}
W_{kX} = (W_{kCX})
\label{theorem2_proof_10.1.1}
\end{eqnarray}

\begin{eqnarray}
M_{CX} = 2^{K h_{XY}}
\label{theorem2_proof_9.1.33}
\end{eqnarray}

where $W_\alpha \in I_{M_\alpha} = \{0, 1, \ldots, M_\alpha - 1\}$. The wiretapper will not know the $W_X$ and $W_Y$ that are secured with keys.

In this case, $R_X$, $R_Y$, $R_{kX}$ and $R_{kY}$ satisfy that

\begin{eqnarray}
\frac{1}{K} [\log M_{kX} + \log M_{kY}] & = & \frac{1}{K} \log M_{CX}	\nonumber
\\ & = & h_{XY} \nonumber \label{num333.1}
\\ & \le & R_{kX} + R_{kY} 
\label{theorem2_proof_13.1.1}
\end{eqnarray}

where \eqref{num333.1} results from \eqref{theorem2_proof_9.1.33}.

The security level thus results:
\begin{eqnarray}
\frac{1}{K} H(X^K, Y^K|W_1, W_2, Z^{\mu}  ) & = & \frac{1}{K} H(X^K, Y^K|W_{X1}, W_{X2}, \nonumber
\\ && W_{CX} \oplus W_{kCX}, \nonumber
\\ && W_{Y1}, W_{Y2},	W_{CY}, Z^{\mu} )			\nonumber
\\ & \geq & \frac{1}{K} \log M_{CX} - \alpha_{CX} - \alpha_{CY} \nonumber
\\ & + & I(X;Y;Z)- \epsilon_0^{'}		\nonumber
\\ & = & h_{XY}- \alpha_{CX} - \alpha_{CY} + I(X;Y;Z) \nonumber
\\ & - & \epsilon_0^{'}  
\label{theorem2_proof_16.1.1}
\end{eqnarray}

where \eqref{theorem2_proof_16.1.1} holds from \eqref{theorem2_proof_9.1.33}.

Therefore ($R_X$, $R_Y$, $R_{kX}$, $R_{kY}$, $h_{XY}$) is admissible from \eqref{theorem2_proof_7.1.1} -  \eqref{theorem2_proof_16.1.1}.

\end{proof}

\begin{proof}[Theorem 2 proof]

In the same way, Suppose that ($R_X$, $R_Y$, $R_{kX}$, $R_{kY}$) $\in$ 
$\mathcal{R}_{2}$ for $h_{X} \le H(X)- \alpha_{CX} - \alpha_{CY} + I(X;Y;Z)$ and $h_Y \le H(Y)- \alpha_{CX} - \alpha_{CY} + I(X;Y;Z)$. Without loss of generality, we assume that $h_X \le h_Y$. Then, from \eqref{theorem3} 
\begin{eqnarray}
&& R_X \geq H(X^K|Y^K)  				\nonumber
\\&& R_Y  \geq H(Y^K|X^K) 				\nonumber
\\&& R_X + R_Y  \geq H(X^K, Y^K)
\label{theorem3_proof_1}
\end{eqnarray}

\begin{eqnarray}
R_{kX} + R_{kY} \geq \text{max}(h_X,h_Y)
\label{theorem3_proof_2}
\end{eqnarray}

Consider the following: $h_X > I(X;Y)$.

\begin{eqnarray}
M_{X1} = \text{min}(2^{K H(X|Y)}, 2^{K(h_Y - I(X;Y))})
\label{theorem3_proof_61}
\end{eqnarray}

\begin{eqnarray}
M_{Y1} = 2^{K (h_Y - I(X;Y))}
\label{theorem3_proof_65}
\end{eqnarray}

The codeword $W_2$ and the key $W_{kY}$ is now defined:

\begin{eqnarray}
W_1 = (W_{X1} \oplus W_{kY1},  W_{X2}, W_{CX} \oplus W_{kCX}) 
\label{theorem3_proof_81}
\end{eqnarray}

\begin{eqnarray}
W_2 = (W_{Y1} \oplus W_{kY1}, W_{Y2}, W_{CY} \oplus W_{kCY})
\label{theorem3_proof_82}
\end{eqnarray}

\begin{eqnarray}
W_{kX} = W_{kCX}
\label{theorem3_proof_83}
\end{eqnarray}

\begin{eqnarray}
W_{kY} = (W_{kY1}, W_{kCY})
\label{theorem3_proof_84}
\end{eqnarray}

In this case, $R_X$, $R_Y$, $R_{kX}$ and $R_{kY}$ satisfy from \eqref{theorem3_proof_61} - \eqref{theorem3_proof_84}, that

\begin{eqnarray}
\frac{1}{K} \log M_X + \frac{1}{K} \log M_Y  & = & \frac{1}{K} (\log M_{X1} + \log M_{X2}  \nonumber
\\ & + &  \log M_{CX}) + \frac{1}{K} (\log M_{Y1}  \nonumber
\\ & + &  \log M_{Y2} +  \log M_{CY}) \nonumber 
\\ & \le & H(X|Y) + H(Y|X) 				\nonumber
\\ & + & I(X;Y) + 3 \epsilon_0 \nonumber
\\ & = & H(X,Y) + 3 \epsilon_0 		\nonumber
\\ & \le & R_X + R_Y + 3 \epsilon_0
\label{theorem3_proof_11}
\end{eqnarray}

\begin{eqnarray}
&& \frac{1}{K} [\log M_{kX} + \log M_{kY}]  \nonumber
\\ & = & \frac{1}{K} [\log M_{CX} + \log M_{CY} + \log M_{Y1}] 	\nonumber	
\\ & \le & I(X;Y) + h_{Y} - I(X;Y) -\epsilon_0  \label{num3.2.1}
\\ & = & h_{Y} - \epsilon_0		\nonumber
\\ & \le & R_{kX} + R_{kY} - \epsilon_0
\label{theorem3_proof_13}
\end{eqnarray}

The security levels thus result:
\begin{eqnarray}
&& \frac{1}{K} H(X^K|W_1, W_2, Z^{\mu})  \nonumber
\\ & = & \frac{1}{K} H(X^K|W_{X1} \oplus W_{kY1}, W_{X2}, W_{CX} \oplus W_{kCX}		\nonumber
\\ && W_{Y1} \oplus W_{kY1}, W_{Y2}, W_{CY} \oplus W_{kCY}, Z^{\mu})			\nonumber
\\ & \ge & \frac{1}{K} H(X^K, Y^K|W_{X1}\oplus W_{kY1}, W_{X2}, W_{Y1} \oplus W_{kY1} 
\\ && W_{Y2}, Z^{\mu}) - \epsilon_0^{''}			\nonumber
\\ & = & \frac{1}{K} H(X^K, Y^K|W_{X2}, W_{Y2}, Z^{\mu}) - \epsilon_0^{''}		\nonumber
\\ & \geq & I(X;Y) + \frac{1}{K} \log M_{X1} - \alpha_{CX} - \alpha_{CY} + I(X;Y;Z)  \nonumber\\
\\ & - &  2\epsilon_0^{'} - \epsilon_0^{''}		\nonumber
\\ & = & I(X;Y)+ \text{min}(2^{K H(X|Y)}, 2^{h_Y - I(X;Y)}) \nonumber
\\ & - & \alpha_{CX} - \alpha_{CY} + I(X;Y;Z) - 2\epsilon_0^{'} - \epsilon_0^{''}	 \nonumber
\\ & \ge & h_{Y} - \alpha_{CX} - \alpha_{CY} + I(X;Y;Z) - 2\epsilon_0^{'} - \epsilon_0^{''}   \nonumber
\\ & \ge & h_{X}
\label{theorem3_proof_16}
\end{eqnarray}

\begin{eqnarray}
\frac{1}{K} H(Y^K|W_1, W_2) & = & \frac{1}{K} H(Y^K|W_{X1} \oplus W_{kX1}, \nonumber
\\ && W_{X2}, W_{CX} \oplus W_{kCX}		\nonumber
\\ && W_{Y1} \oplus W_{kY1}, W_{Y2} \nonumber
\\ && W_{CY} \oplus W_{kCY}, Z^{\mu})			\nonumber
\\ & \ge & \frac{1}{K} \log M_{Y1} + I(X;Y) - \alpha_{CX} \nonumber
\\ & - & \alpha_{CY} + I(X;Y;Z) -\epsilon_0^{'} \nonumber
\\ & = & I(X;Y) + \text{min} (H(X|Y), h_{Y} - I(X;Y)) \nonumber
\\ & - & \alpha_{CX} - \alpha_{CY} + I(X;Y;Z) -\epsilon_0^{'}			\label{num5.44}
\\ & \ge & h_{Y} - \epsilon^{'}_0
\label{theorem3_proof_161}
\end{eqnarray}

where \eqref{num5.44} comes from \eqref{theorem3_proof_65}.

Therefore ($R_X$, $R_Y$, $R_{kX}$, $R_{kY}$, $h_{X}$, $h_{Y}$) is admissible from \eqref{theorem3_proof_11} -  \eqref{theorem3_proof_161}.

Next the case where $h_X \le I(X;Y)$ is considered. If $h_Y > I(X;Y)$ the following results.
The codewords $W_1$ and $W_2$ and their keys $W_{kX}$ and $W_{kY}$ are now defined:

\begin{eqnarray}
W_1 = (W_{X1},  W_{X2}, W_{CX} \oplus W_{kCX}) 
\label{theorem3_proof_7.1.1}
\end{eqnarray}

\begin{eqnarray}
W_2 = (W_{Y1} \oplus W_{kY1}, W_{Y2}, W_{CY} \oplus W_{kCY})
\label{theorem3_proof_8.1.1}
\end{eqnarray}

\begin{eqnarray}
W_{kX} = (W_{kCX})
\label{theorem3_proof_10.1.1}
\end{eqnarray}

\begin{eqnarray}
W_{kY} = (W_{kY1}, W_{kCY})
\label{theorem3_proof_10.1.2}
\end{eqnarray}

\begin{eqnarray}
M_{Y1} = 2^{K(h_{Y} - I(X;Y)})
\label{theorem3_proof_9.1.333}
\end{eqnarray}

where $W_\alpha \in I_{M_\alpha} = \{0, 1, \ldots, M_\alpha - 1\}$. The wiretapper will not know the $W_X$ and $W_Y$ that are secured with keys.

In this case, $R_X$, $R_Y$, $R_{kX}$ and $R_{kY}$ satisfy that

\begin{eqnarray}
\frac{1}{K} [\log M_{kX} + \log M_{kY}] & = & \frac{1}{K} [\log M_{CX} + \log M_{Y1} + \log M_{CY}]	\nonumber
\\ & \le & I(X;Y) + \frac{1}{K} \log M_{Y1} -\epsilon_0  \nonumber
\\ & = & I(X;Y) + h_{Y} - I(X;Y) - \epsilon_0  \label{num333.11}
\\ & = & h_{Y} - \epsilon_0 \nonumber
\\ & \le & R_{kX} + R_{kY} + \epsilon_0
\label{theorem3_proof_13.1.1}
\end{eqnarray}

where \eqref{num333.11} results from \eqref{theorem3_proof_9.1.333}.

The security levels thus result:
\begin{eqnarray}
&& \frac{1}{K} H(X^K|W_1, W_2) \nonumber
\\ & = &  \frac{1}{K} H(X^K|W_{X1}, W_{X2} \nonumber
\\ && W_{CX} \oplus W_{kCX}, W_{Y1} \oplus W_{kY1}, W_{Y2},		\nonumber
\\ && W_{CY} \oplus W_{kCY}, Z^{\mu})			\nonumber
\\ & \geq & I(X;Y)  - \alpha_{CX} - \alpha_{CY} + I(X;Y;Z) - \epsilon_0^{'}	
\\ & = & I(X;Y) - \alpha_{CX} - \alpha_{CY} + I(X;Y;Z)  - \epsilon^{'}_0	\label{num5.1.1}
\\ & \geq & h_X - \epsilon^{'}_0
\label{theorem3_proof_16.1.1}
\end{eqnarray}

where \eqref{num5.1.1} results from \eqref{theorem3_proof_9.1.333}.

\begin{eqnarray}
&&\frac{1}{K} H(Y^K|W_1, W_2) \nonumber
\\ & = & \frac{1}{K} H(Y^K|W_{X1}, W_{X2} \nonumber
\\ && W_{CX} \oplus W_{kCX}, W_{Y1} \oplus W_{kY1}, 		\nonumber
\\ && W_{Y2}, W_{CY} \oplus W_{kCY}, Z^{\mu})			\nonumber
\\ & \geq & I(X;Y) + \frac{1}{K} \log M_{Y1} \nonumber
\\ & - & \alpha_{CX} - \alpha_{CY} + I(X;Y;Z) - \epsilon_0	
\\ & = & I(X;Y) + h_{Y} - I(X;Y) \nonumber
\\ & - & \alpha_{CX} - \alpha_{CY} + I(X;Y;Z)  - \epsilon^{'}_0	\label{num5.1.11}
\\ & \ge & h_{Y} - \epsilon^{'}_0
\label{theorem3_proof_16.1.1}
\end{eqnarray}

where \eqref{num5.1.11} holds from \eqref{theorem3_proof_9.1.333}.

Next the case where $h_Y \le I(X;Y)$ is considered. 
The codewords $W_1$ and $W_2$ and their keys $W_{kX}$ and $W_{kY}$ are now defined:

\begin{eqnarray}
W_1 = (W_{X1},  W_{X2}, W_{CX} \oplus W_{kCX}) 
\label{theorem3_proof_7.1.1.1}
\end{eqnarray}

\begin{eqnarray}
W_2 = (W_{Y1}, W_{Y2}, W_{CY})
\label{theorem3_proof_8.1.1.1}
\end{eqnarray}

\begin{eqnarray}
W_{kX} = W_{kCX}
\label{theorem3_proof_10.1.1.1}
\end{eqnarray}

\begin{eqnarray}
M_{CX} = 2^{K h_Y}
\label{theorem3_proof_9.1.333.1}
\end{eqnarray}

where $W_\alpha \in I_{M_\alpha} = \{0, 1, \ldots, M_\alpha - 1\}$. The wiretapper will not know the $W_X$ and $W_Y$ that are secured with keys.

In this case, $R_X$, $R_Y$, $R_{kX}$ and $R_{kY}$ satisfy that

\begin{eqnarray}
\frac{1}{K} [\log M_{kX} + \log M_{kY}] & = & \frac{1}{K} \log M_{CX}	\nonumber
\\ & = & h_Y \label{num333.11.1}
\\ & \le & R_{kX} + R_{kY}
\label{theorem3_proof_13.1.1}
\end{eqnarray}

where \eqref{num333.11.1} results from \eqref{theorem3_proof_9.1.333.1}.

The security levels thus result:
\begin{eqnarray}
&& \frac{1}{K} H(X^K|W_1, W_2, Z^{\mu}) \nonumber
\\ & = & \frac{1}{K} H(X^K|W_{X1}, W_{X2} \nonumber
\\ && W_{CX} \oplus W_{kCX}, W_{Y1}, W_{Y2},		\nonumber
\\ && W_{CY}, Z^{\mu})			\nonumber
\\ & \geq & h_Y  -  \alpha_{CX} - \alpha_{CY} + I(X;Y;Z)  - \epsilon_0^{'}	\label{num5.1.1.1}
\\ & \geq & h_X - \epsilon^{'}_0
\label{theorem2_proof_16.1.1}
\end{eqnarray}

where \eqref{num5.1.1.1} results from \eqref{theorem3_proof_9.1.333.1}.

\begin{eqnarray}
\frac{1}{K} H(Y^K|W_1, W_2, Z^{\mu}) & = & \frac{1}{K} H(Y^K|W_{X1}, W_{X2} \nonumber
\\ && W_{CX} \oplus W_{kCX}, W_{Y1}, W_{Y2},		\nonumber
\\ && W_{CY}, Z^{\mu})			\nonumber
\\ & \geq & \frac{1}{K} \log M_{CY} - \alpha_{CX} - \alpha_{CY}  \nonumber
\\ & + & I(X;Y;Z) - \epsilon_0^{'}  \label{num5.1.11.1}
\\ & \geq & h_Y -  \alpha_{CX} - \alpha_{CY} + I(X;Y;Z) 	
\label{theorem3_proof_16.1.1.1}
\end{eqnarray}

where \eqref{num5.1.11.1} holds from \eqref{theorem3_proof_9.1.333.1}.

Therefore ($R_X$, $R_Y$, $R_{kX}$, $R_{kY}$, $h_{X}$, $h_{Y}$) is admissible for $\text{min}(h_X, h_Y)$ \eqref{theorem3_proof_7.1.1.1} -  \eqref{theorem3_proof_16.1.1.1}.
\end{proof}

\subsection{Converse parts}
From Slepian-Wolf's theorem we know that the channel rate must satisfy $R_X \geq H(X|Y)$, $R_Y \geq H(Y|X)$ and $R_X + R_Y \geq H(X,Y)$ to achieve a low error probability when decoding.
Hence, only the key rates are considered in this subsection. 
\\
\textit{Converse part of Theorem 1:}
\begin{eqnarray}
R_{kX} & \geq & \frac{1}{K} \log M_{kX} - \epsilon		\nonumber
\\ & \geq & \frac{1}{K} H(W_{kX}) - \epsilon			\nonumber
\\ & \geq & \frac{1}{K} H(W_{kX}|W) - \epsilon			\nonumber
\\ & = & \frac{1}{K} [H(W_{kX}) - I(W_{kX}; W)] - \epsilon		\nonumber
\\ & = & \frac{1}{K} H(W_{kX}|X^K, Y^K, W) + I(W_{kX}; W) 		\nonumber
\\ & + & I(W_{kX};X|Y, W) + I(X, Y, W_{kX}|W) 		\nonumber
\\ & + & I(Y, W_{kX}|X, W) - I(W_{kX}; W) - \epsilon		\nonumber
\\ & = & \frac{1}{K} [H(X^K, Y^K|W) - H(X^K,Y^K|W, W_{kX})] - \epsilon		\nonumber
\\ & \geq & h_{XY} - \frac{1}{K} H(X^K,Y^K|W, W_{kX})  - \epsilon  \label{conv_1}
\\ & = & h_{XY} - \frac{1}{K} H(Y^K|X^K) - \alpha_{CX} - \alpha_{CY} \nonumber
\\ & + & I(X;Y;Z) - \epsilon  	- \epsilon_0^{''}	\nonumber
\\ & \geq & h_{XY} - \alpha_{CX} - \alpha_{CY} + I(X;Y;Z) - \epsilon - \epsilon_0^{''}
\end{eqnarray}

where $W = (W_1, W_2, Z^{\mu})$ are the wiretapped portions, \eqref{conv_1} results from equation \eqref{cond8.1}. Here, we consider the extremes of $H(Y|X)$ and $H(W_Y)$ in order to determine the limit for $R_{kX}$. When this quantity is minimum then we are able to achieve the maximum bound of $h_{XY}$.

\begin{eqnarray}
R_{kY} & \geq & \frac{1}{K} \log M_{kY} - \epsilon		\nonumber
\\ & \geq & \frac{1}{K} H(W_{kY}) - \epsilon			\nonumber
\\ & \geq & \frac{1}{K} H(W_{kY|W}) - \epsilon		\nonumber
\\ & = & \frac{1}{K} [H(W_{kY}) - I(W_{kY}; W) - \epsilon		\nonumber
\\ & = & \frac{1}{K} H(W_{kY}|X, Y, W) + I(W_{kY}; W) 		\nonumber
\\ & + & I(W_{kY};X|Y, W) + I(X, Y, W_{kY}|W) 				\nonumber
\\ & + & I(Y, W_{kY}|X, W) - I(W_{kY}; W)] - \epsilon			\nonumber
\\ & = & \frac{1}{K} [H(X^K, Y^K|W) - H(X^K,Y^K|W, W_{kY})] - \epsilon		\nonumber
\\ & \geq & h_{XY} - \frac{1}{K} H(X^K,Y^K|W, W_{kY})  - \epsilon  \label{conv_2}
\\ & = & h_{XY} - \frac{1}{K}H(X^K|Y^K)- \alpha_{CX} - \alpha_{CY}  \nonumber
\\ & + & I(X;Y;Z) - \epsilon  - \epsilon_0^{''} 	\nonumber
\\ & \geq & h_{XY} - \alpha_{CX} - \alpha_{CY} + I(X;Y;Z)  - \epsilon - \epsilon_0^{''}
\end{eqnarray}

where \eqref{conv_2} results from equation \eqref{cond8.1}. Here, we consider the extremes of $H(V_{CX})$ in order to determine the limit for $R_{kY}$. When this quantity is minimum then we are able to achieve the maximum bound of $h_{XY}$.

\textit{Converse part of Theorem 2:}
\begin{eqnarray}
R_{kX} & \geq & \frac{1}{K} \log M_{kX} - \epsilon		\nonumber
\\ & \geq & \frac{1}{K} H(W_{kX}) - \epsilon		\nonumber
\\ & \geq & \frac{1}{K} H(W_{kX}|W) - \epsilon		\nonumber
\\ & = & \frac{1}{K} [H(W_{kX}) - I(W_{kX}; W)] - \epsilon		\nonumber
\\ & = & \frac{1}{K} H((W_{kX}|X^K, W) + I(W_{kX}; W) 		\nonumber
\\ & + & I(X, W_{kX}|W) - I(W_{kX}; W) - \epsilon		\nonumber
\\ & \geq & \frac{1}{K} I(X^K, W_{kX}|W) - \epsilon		\nonumber
\\ & = & \frac{1}{K} [H(X^K|W) - H(X^K|W, W_{kX})] - \epsilon	\nonumber
\\ & \geq & h_{X} - H(W_{CY}) - \alpha_{CX} - \alpha_{CY} + I(X;Y;Z)  \nonumber
\\ & - & \epsilon - \epsilon_0^{''}  \label{conv_3}
\\ & \geq & h_{X}   - \alpha_{CX} - \alpha_{CY} + I(X;Y;Z)  -  \epsilon - \epsilon_0^{''}
\end{eqnarray}

where $W= (W_1, W_2, Z^{\mu})$, \eqref{conv_3} results from \eqref{cond7}. The consideration here is that $H(Y^{K_2})$ represents the preexisting information known to an eavesdropper as an extreme case scenario.
Here, we can also consider the extremes of $H(W_{CY})$ in order to determine the limit for $R_{kX}$. When this quantity is minimum then we are able to achieve the maximum bound of $h_{X}$.

\begin{eqnarray}
R_{kY} & \geq & \frac{1}{K} \log M_{kY} - \epsilon	\nonumber
\\ & \geq & \frac{1}{K} H(W_{kY}) - \epsilon		\nonumber
\\ & \geq & \frac{1}{K} H(W_{kY}|W) - \epsilon		\nonumber
\\ & = & \frac{1}{K} [H(W_{kY}) - I(W_{kY}; W)] - \epsilon		\nonumber
\\ & = & \frac{1}{K} H(W_{kY}|Y^K, W) + I(W_{kY}; W) 		\nonumber
\\ & + & I(X, W_{kY}|W) - I(W_{kY}; W) - \epsilon		\nonumber
\\ & \geq & \frac{1}{K} I(Y^K, W_{kY}|W) - \epsilon				\nonumber
\\ & = & \frac{1}{K} [H(Y^K|W) - H(Y^K|W, W_{kY})] - \epsilon		\nonumber
\\ & \geq & h_{Y} - H(W_{CX}) - \alpha_{CX} - \alpha_{CY} + I(X;Y;Z) \nonumber
\\ & - & \epsilon - \epsilon_0^{''} \label{conv_4}
\\ & \geq & h_{Y} - \alpha_{CX} - \alpha_{CY} + I(X;Y;Z) - \epsilon - \epsilon_0^{''}
\end{eqnarray}

where \eqref{conv_4} results from \eqref{cond8}. The same consideration as above for $ H(Z^{\mu})$ is presented here. Here, we consider the extremes of $H(W_{CX})$ in order to determine the limit for $R_{kY}$. When this quantity is minimum then we are able to achieve the maximum bound of $h_{Y}$.

\bibliographystyle{IEEEtran}	% (uses file "IEEEtran.bst")
\bibliography{bib}

\end{document}